# Non-traditional data in pandemic preparedness and response: identifying and addressing first and last-mile challenges

Mattia Mazzoli[1*], Irma Varela-Lasheras[2*], Sonia Namorado[2,3], Constantino Pereira Caetano[2], Andreia Leite[2,3], Lisa Hermans[4,5], Niel Hens[4,5], Polen Türkmen[6,1], Kyriaki Kalimeri[1], Leo Ferres[1,7], Ciro Cattuto[1], Daniela Paolotti[1], Stefaan Verhulst[1,6,8,9]

[1] ISI Foundation, Turin, Italy
[2] Department of Epidemiology, National Institute of Health Doctor Ricardo Jorge, Lisbon, Portugal
[3] NOVA National School of Public Health, Public Health Research Centre, Comprehensive Health Research Center, CHRC, LA-REAL, CCAL, NOVA University Lisbon, Lisbon, Portugal
[4] Interuniversity Institute of Biostatistics and statistical Bioinformatics, Data Science Institute, Hasselt University, Hasselt, Belgium
[5] Centre for Health Economics Research and Modelling Infectious Diseases, Vaccine and Infectious Disease Institute, University of Antwerp, Antwerp, Belgium
[6] The Data Tank
[7] Data Science Institute, Universidad del Desarrollo, Santiago de Chile, Chile
[8] The GovLab, New York University, New York, USA
[9] Interuniversity MicroElectronics Center - IMEC - SMIT, Vrije Universiteit Brussel, Brussels, Belgium

* = co-first authorship

## Abstract

**Background**
The COVID-19 pandemic served as an important test case of complementing traditional public health data with non-traditional data such as mobility traces, social media activity, and wearables data to inform real-time decision-making.

**Objective**
Drawing on an expert workshop and a targeted survey of European modelers, this article assesses the promise and persistent limitations of such data in pandemic preparedness and response. We distinguish between "first-mile" challenges (obstacles to accessing and harmonizing data) and "last-mile" challenges (difficulties in translating insights into actionable policy interventions).

**Methods**
The expert workshop convened in March 2024 in Brussels brought together 50 participants including public health professionals, data scientists, policymakers, and industry leaders to reflect on lessons learned and define strategies for better integration of non-traditional data into epidemic modeling and policy making. The accompanying survey, gathering experiences from 29 modelers, offers empirical evidence of the barriers faced by modelers during COVID-19 and highlights areas where key data was unavailable or underutilized.

**Results**
Our findings reveal ongoing issues with data access, quality, and interoperability, as well as institutional and cognitive barriers to evidence-based decision-making. Approximately 66% of all datasets suffered at least one access problem, with data sharing reluctance for non-traditional sources being double that of traditional data (30% vs 15%). Only 10% of respondents reported they could use all the data they needed. These limitations included timeliness and granularity of data, issues with linkage, comparability, and biases.

To overcome these hurdles, we propose a set of enabling mechanisms, including data inventories, standardization protocols, simulation exercises, data stewardship roles, and data collaboratives. For first-mile challenges, solutions focus on technical and legal frameworks for data access. For last-mile challenges, we recommend fusion centers, decision accelerator labs, and networks of scientific ambassadors to bridge the gap between analysis and action.

**Conclusions**

We argue that realizing the full value of non-traditional data requires a sustained investment in institutional readiness, cross-sectoral collaboration, and a shift toward a culture of data solidarity. Grounded in the lessons of COVID-19, the article can be used to design a roadmap for using non-traditional data to confront a broader array of public health emergencies, from climate shocks to humanitarian crises.

**Keywords**

non-traditional data; pandemic preparedness; pandemic response

**Introduction**

Throughout the COVID-19 pandemic, decision-makers around the world looked for timely and quality information to inform their response efforts. Yet, much of the data traditionally used for public health (e.g., public health-based surveillance data, healthcare-based data, clinical trials) were not available fast enough or at the scale or the coverage needed for a crisis that extended beyond national borders and required a multi-sector, timely response. As a result, many in the scientific community turned to new, non-traditional data sources and cross-sectoral collaborations to fill these gaps and accelerate knowledge generation to support decision-making. We acknowledge the importance of mathematical, statistical and AI tools to interpret data and provide support for evidence-based policy making. However, here we explicitly focus on the two endings of the process, i.e., availability and retrieval of critical data and their power to inform public health decisions.

Non-traditional data (NTD) is often defined as: "repurposed data that can be digitally captured (e.g., mobile phone and financial data), mediated (e.g., social media and digital traces), or observed (e.g., satellite imagery)"[1–3], often using new and privately held technology[3]. As such, it is an umbrella term for information not originally collected for public health but repurposed for that aim (see **Table 1**). Since such feeds are often generated continuously and at a population scale, they offer more granular and more real-time data, helping public health responders, governments, and civil society organizations to respond more effectively.

|  | Traditional data | NTD |
|---|---|---|
| **Purpose of data collection** | Population health | Business optimization, transport planning, environmental control, algorithm optimization, social behavior analysis, flu-like illness monitoring |
| **Collection method** | Analog and digital | Digital |

| Coverage type | Groups targeted through recruitment, healthcare seeking and notifiable diseases | Groups defined by digital service use and technology adoption - often scalable to the population level |
|---|---|---|
| Structure | Structured | Often unstructured |
| Cause of data collection | Population health | Repurposed for population health |
| Example | Cohort studies, clinical trials, traditional contact tracing, consultation rates, notifiable diseases | XDR mobility data, participatory surveillance, social media data, digital contact tracing |

**Table 1. Properties of traditional and non-traditional data**

The potential of non-traditional data to inform the response to public health emergencies has already long been recognized. For example, during the 2014–2016 West Africa Ebola outbreak, anonymized mobile-phone call records were used to map population movements and shape travel-restriction policies in Sierra Leone[4]. Similarly, Google search queries have been used to predict and offer early warnings on diseases, including Zika[5]. The use of non-traditional data has been an important aspect of many of these efforts. Its use was further highlighted during the COVID-19 crisis, illustrating how such a rapid and granular data source can be employed to inform the response to a public health crisis.

Typical non-traditional data sources include aggregated and anonymized mobile location traces, information on retail purchases, social media posts, and data sourced from wearables, for instance, on heart rate or temperature[3]. NTD have been classified in four main types: health, social mixing, economic, and sentiment data (see **Figure 1**). Applications to pandemic response are exemplified below.

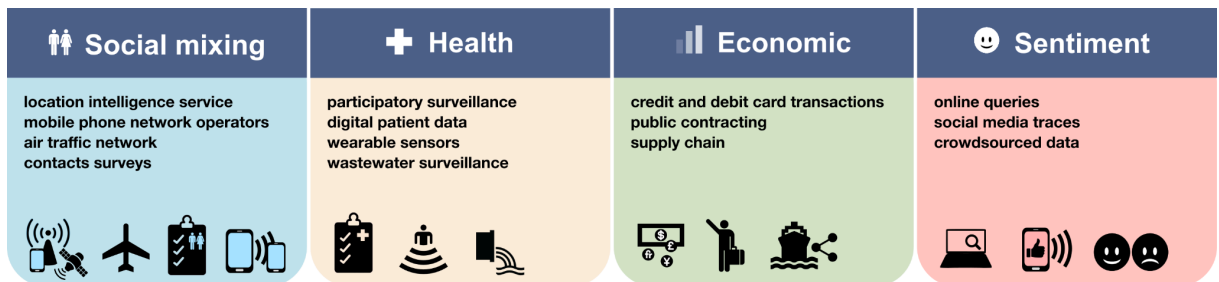

**Figure 1. Non-Traditional Data types and sources**

**Health**: At the core of pandemic response stands the monitoring of population health. At the beginning of the pandemic, researchers quickly repurposed pre-existing crowdsourcing platforms to monitor COVID-19 circulation and related behavior[6]. The participatory surveillance[7] platform InfluenzaNet[8] has collected flu-like symptoms and health-related behavior from volunteers across Europe since 2003 to analyze country-level incidence trends on a weekly basis[9]. While volunteers report symptoms to their respective country's unique platform, all national platforms within the InfluenzaNet umbrella adopt standardized approaches to reporting, facilitating cross-country analysis. Influenzanet expanded its metrics infrastructure in 2020 to include COVID-19 data. Similar efforts are found in the US with OutbreaksNearMe[10], in South Africa with CoughWatch[11], and in Australia and New Zealand with Flutracking[12]. In many cases, symptom reports are collected through mobile applications, as in the case of a mobile application[13] launched by King's College London and ZOE, a science and nutrition company. These data helped public health officials in the United Kingdom to identify early indicators of illness in the emergent stages of the pandemic. Smart thermometers and other wearable sensors were a precious data source for analysing trends of fever in the population. Wearable sensor data[14] collected by Kinsa's network of smart thermometers[15] has been employed to analyse fever patterns and develop outbreak forecasting maps throughout several countries. Thermometer data were already used in 2012 to forecast influenza outbreaks and were repurposed for COVID-19 during the onset of the pandemic. A wide range of stakeholders used these data, including pharmacies, schools, and other decision makers. In Germany, wearable technology data collected by Thryve, in partnership with the Robert Koch Institute, created a data donation platform[16] for improved fever detection research in the spring of 2020. Wastewater-based epidemiology (WBE), which is based on sampling of sewage waters, proved to be a powerful tool during the pandemic to achieve a more granular and unbiased method to monitor disease circulation[17,18]. WBE can detect viral traces of SARS-CoV-2 among other biological and environmental contaminants, and reflect transmission dynamics in untested communities[19], allowing for earlier detection of outbreaks. In Hong Kong, wastewater surveillance efforts found the first evidence of the spread of the Delta variant and informed public health interventions[20].

**Social mixing**: On a geographic scale, mobility patterns encode mixing of residents of different locations, allowing modelers to estimate the spatial spread of the disease and suggest spatially targeted interventions. On a population level, contact data encode interactions at risk for contagion, allowing modelers to study disease transmission across demographic groups and suggest settings-specific interventions. Mobility records were published regularly after the early days of the pandemic through *Data for Good*[21] programs at diverse spatial scales and temporal resolutions. Google, for instance, regularly published human mobility data through its Community Mobility Reports[22]. These data regarded worldwide regions and included trends around commonly visited location types (e.g., retail, workplace, residential) that, crossed with census data, helped inform epidemic modeling to support decision makers. Social media companies' *Data for Good* programs, such as Meta[23] and Cuebiq[24], assisted epidemic modelers in providing critical insights for public health measures and population studies. Telcos[25] played an important role in tackling the pandemic in their respective countries, helping modelers and public health practitioners monitor and understand the heterogeneous population response to restrictions, as in the case of Telefonica Chile[26] and Orange France[27]. These data helped decision-makers assess the effectiveness of their policies, monitor population compliance, and tailor public health response. Beyond mobility patterns, social mixing data were typically collected from traditional contact surveys and diaries, which did not scale well at the national level and did not offer longitudinal or real-time follow-up of human behavior. Following pre-pandemic efforts such as POLYMOD[28], online contact surveys re-emerged in the pandemic's early days thanks to new crowdsourcing tools at the European level, such as CoMix[29], an initiative funded by the European Commission (EC) that quickly gathered public mixing and sentiment data during the COVID-19 pandemic. The weekly survey started in March 2020 in Belgium, the Netherlands, and the United Kingdom before expanding to 17 additional countries across Europe. Marketing firm Ipsos MORI recruited participants via social media ads and email campaigns to crowdsource reported social contacts, assisting with national policy evaluation.

**Economic:** The pandemic shifted consumers' needs and habits, impacting local businesses' revenues and opening new market opportunities. Economic NTD helped policymakers monitor the impact of the pandemic on local businesses and the population's spending habits. Data on electronic transactions[30] inform us about the time, location, and purpose of people's purchases, proxying their economic wealth. Electronic transaction data from credit cards were used to measure the reduction in revenue[31] of economic activities during the COVID-19 pandemic.
Supply chain data allow to monitor the status of the production and distribution of goods, providing logistics organisations with important insights to optimise the goods pipeline. Using maritime traffic data collected via a global network of Automatic Identification System (AIS) receivers, the Marine Traffic Research Lab studied the impact of the pandemic on the shipping industry[32]. Open contracting records refer to records of public contracts from administrations. They are used in many countries to increase transparency and contrast corruption in public spending. In Ecuador, the National Public Procurement Service launched a public-facing dashboard of emergency procurement contracts during COVID-19[33].

**Sentiment**: While aiming at controlling disease circulation, interventions had an impact on population mental distress and were partially received by citizens, giving rise to misinformation exposure and loss of adherence to policies over time. Sentiment NTD allowed researchers and organisations to monitor people's perceptions, mental health, and exposure to misinformation. Online search queries and social media data were valuable tools to understand the impact of online behavior on the pandemic progression and monitor the population's perception of restrictions. Google search trends and Tweets were used by German researchers to create a future-oriented early alert system[34] that could be used to build phenomenological and mechanistic models to forecast cases and hospitalization trends. Tweets before and during the pandemic were analysed by researchers in Italy to analyse the potential health threat driven by online misinformation[35] related to the pandemic. The same data was used by researchers in the US to integrate public health data with social media data, developing an early alert system for outbreak detection[36]. However, sentiment data do not stop at online discourse on social media; the worldwide Covid-19 trends and impact survey[37] conducted by Meta collected important information, among others, on the population's mental health. Over the course of the pandemic, respondents transmitted self-reported feelings of anxiety and depression, for the first time collecting important insights on the population's mental distress and risk perception during a pandemic. While CoMix[29] collected both self-reported behavioral data and sentiment from participants, other efforts aimed to incorporate a wider variety of data streams related to people's perception of the pandemic. For example, six European public health agencies collaborated during the pandemic to create PANDEM-Source[38], an IT surveillance tool that integrates both traditional case counts and non-traditional social media reports for pandemic surveillance. Results feed into PANDEM-2[39], a real-time data dashboard for decision-makers and health experts.

Non-traditional data can help fill critical gaps in traditional data during public health emergencies by increasing the speed and volume of data collection. When combined with traditional data, non-traditional data can provide a more comprehensive understanding of the problem and support rapid decision-making. However, alongside the promise, COVID-19 also revealed challenges involved in the use of both traditional and non-traditional data. The early months of the COVID-19 pandemic were characterized by ad-hoc collaborations, which revolved around the use of non-traditional data sources: mobility dashboards built overnight[40], wastewater pilot studies[20], and hastily negotiated data-sharing agreements[41].
A noteworthy example of ad-hoc collaboration between public bodies and the private sector was the European Commission's collaboration with European Mobile Network Operators (MNOs)[42]. Within just a few months, 17 MNOs across 22 EU states, and Norway provided access to mobile phone data with the aim of understanding the geographical spread of the disease and assessing pandemic interventions. In addition to European-level initiatives, many sub-national regions, such as the Valencian region of Spain[43], combined data from mobile operators with data from Facebook, Google, and government ministries to evaluate the impact of mobility restrictions on COVID-19 spread. Challenges that emerged from these pioneering initiatives included data harmonization and privacy guardrails.

Even after the pandemic, we can find virtuous examples of public-private collaborations allowing non-traditional data collection aimed at making data publicly accessible for research. One institutional example is represented by the initiative of the Transport Ministry of Spain[44], regularly making publicly available origin-destination matrices[45] of mobility flows at high temporal and spatial resolution in Spain since early 2020. This ongoing initiative is a collaboration between the Ministry, Orange Spain, and Nommon, and provides a curated and post-stratified, ready-to-use dataset of human mobility stratified by multiple socio-demographics. On a more bottom-up side, the community of researchers is also responding to the growing lack of non-traditional data accessibility. This is the example of R2M2P2 Consortium: Readying Regional Mobility data for Modelling Pandemic Preparedness[46], led by the Erasmus Medical Center in collaboration with Dutch universities and the national public health agency RIVM. The aim of this effort is to identify useful mobility data sources and set up a pipeline providing usable and accessible mobility data to inform transmission models to respond to key public health questions on the spread of respiratory infections in the Dutch population.

These innovations helped highlight how diverse data streams, when responsibly harnessed, can quicken detection times and potentially reduce mortality and morbidity. These experiences served as an important test case for the integration of these new sets of data into epidemic modeling and decision-making for pandemic response. However, they also exposed deep structural weaknesses, such as fragmented governance, limited interoperability, and a chronic skills gap. Questions remain around how these data can best be used in a pandemic crisis. Namely, regarding its limitations, and the investment needed to replicate efforts from COVID-19 in future public health emergencies. A better understanding of these challenges may provide insights and lessons for future efforts to combat public health emergencies.

This article is based on the structured discussions held during an expert workshop on the use of non-traditional data to counteract pandemics conducted in March 2024 in Brussels, along with the results of a survey on data readiness and availability disseminated among the European modeling community conducted the year before. The workshop convened public health professionals, data scientists, policymakers, industry leaders, and other stakeholders to reflect on lessons learned and define strategies for better integration of non-traditional data into epidemic modeling and policy making. The accompanying survey offers empirical evidence of the barriers faced by modelers during COVID-19 and highlights areas where key data was unavailable or underutilized. Together, these two sources form the basis for ten key lessons learnt and proposed actions about the role of non-traditional data in pandemic preparedness and response, which we outline below.

Based on the experts' opinions and the conducted survey, we draw a set of recommendations that establish a framework for the use of non-traditional data in ongoing and future public health and other crises. In particular, with hindsight, it is now clear that two main categories of obstacles emerged: "first-mile" and "last-mile" challenges. First-mile challenges refer to the difficulties of rapidly identifying and accessing relevant data when a crisis begins; these include legal constraints, unclear stewardship, and difficulty in the application of the FAIR[47] principles, i.e., lack of interoperability and, critically, unavailability of data. Last-mile challenges are more nuanced and often less recognized, but are nonetheless critical. They concern the difficulty of translating complex data into actionable insights that decision-makers can understand, trust and apply. Together, these challenges represent a formidable and persistent obstacle to the use of non-traditional data in public health emergencies. To move beyond improvisation during a crisis, we argue for a more proactive, institutionalized, and scalable approach to utilising non-traditional data streams before, during, and after public health emergencies.

**Methods**

**Survey on data availability and data needs**
In 2024 we conducted a survey[48] and circulated it to 29 experts (individuals and groups) from academia, government agencies, and the private sector to better understand data availability, use and unmet needs during the COVID-19 pandemic, primarily in the European context. Specifically, the survey aimed to gather which types of data respondents drew on during COVID-19, obstacles and challenges they faced, and whether and how these data was used to shape policy.

The survey was divided into 3 sections. The first one aimed to collect research questions and modeling approaches used by researchers participating in the survey. The second section aimed to identify (1) what data was available and has been used during the pandemic; (2) how data was accessed and what problems were encountered in accessing the data; and (3) what quality problems were encountered when using the data. Finally, the third section aimed to understand what data was needed but not used by these research groups and whether these unmet needs were the result of data unavailability or the result of data quality and access issues. Throughout the survey, the data was categorized in traditional epidemiological data (e.g. tests, hospitalizations, deaths, vaccination, etc.) and non-traditional epidemiological data, i.e. data types not collected with epidemiological purposes and/or relatively new data types (e.g. mobility, contacts, wastewater data,etc.), so that in addition to each individual data type, we can have a broader view on traditional vs. non-traditional data. The survey was disseminated among modelers in the ESCAPE project as well as among the modeling community in Europe (through, for instance, the ECDC Modeling Hubs mailing list).
In total, 29 research groups/researchers responded to the survey.

**Workshop on first and last-mile challenges for data needs**
On March 22nd 2024, we held a workshop at the University Foundation in Brussels to better understand the role (including limitations) of non-traditional data. Attendance of the Workshop was upon invitation. We assembled a high-level expert panel reflecting the workshop's interdisciplinary goals, bringing together approximately 50 participants from diverse sectors:

1.  **Private sector**: Representatives from industries such as telecom companies, which typically own and provide relevant non-traditional data.
2.  **Academic sector**: Leading European scientists and project partners who played pivotal roles during the pandemic response.
3.  **Public Health representatives**: Officials from national public health institutes (e.g., Portugal, Switzerland) and European institutions (e.g., Hera).

Attendees were selected based on their expertise and background. In the figure below, a summary figure of the provenance of the invited participants can be found:

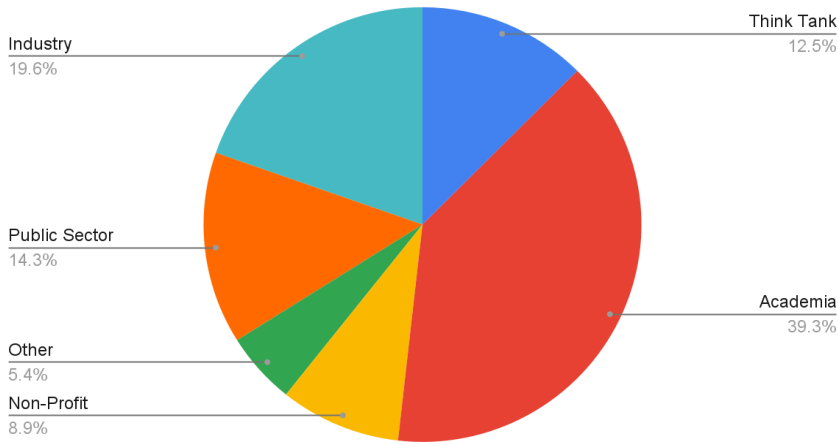

**Figure 2. Workshop participating organizations.**

The workshop's specific objectives included:

- Advancing the dialogue on leveraging non-traditional data for pandemic preparedness.
- Drawing insights from experts and practitioners in relevant fields.
- Building meaningful collaborations to enhance pandemic response efforts.
- Co-creating high-level recommendations and a roadmap for European decision-makers.

Discussions and activities were structured around two main themes: "Data for Preparedness" and "Data for Pandemic Response." Participants were divided into working groups to explore these topics through three key sessions

1. **First-mile challenges**: Addressing readiness to access non-traditional data in case of a pandemic. Questions included: Are we prepared to access and re-use non-traditional data for preparedness and response? What challenges remain from the previous pandemic?
2. **Prototyping first-mile solutions**: This session focused on refining and defining challenges, gathering quick feedback, identifying areas with the greatest impact or feasibility, and proposing concrete actions and recommendations.
3. **Last-mile challenges**: Exploring how to ensure data insights are translated into actionable decision-making. Discussions centered on identifying key decision-makers, facilitating partnerships between data, scientific, and policy communities, and fostering effective communication.

The final plenary session united all participants to review and discuss the solutions proposed during the previous sessions.

**Results**

**Pandemic response as a hybrid enterprise**

The survey responses suggest that non-traditional data sources have moved from fringe to mainstream. For instance, 90% of modelers reported using at least one non-traditional stream alongside more conventional sources. Among non-traditional data types, mobility was dominant (72% of respondents), alongside other non-traditional sources such as socio-economic and demographic, contact, and wastewater data (see **Fig.3**). An equally striking aspect of the survey finding was the direct relevance of such data to inform the response to the pandemic. 79% of the respondents stated that they used data to inform public health policies, with the largest number of these (83%) stating that they used mathematical models to translate data into insights used to inform policy making, followed by statistical models (48% of respondents).

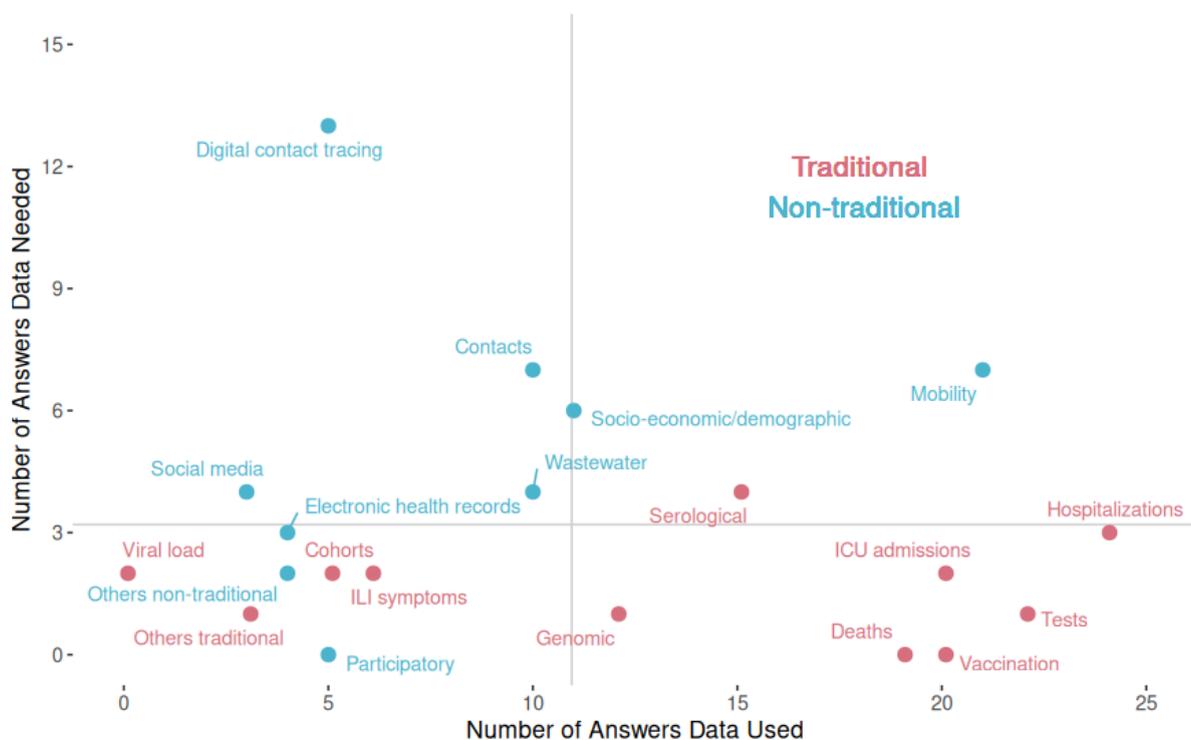

**Figure 3**. **Comparison of the use (x-axis) and the need (y-axis) of the different types of data reported by respondents.** Vertical and horizontal grey lines represent the mean number of answers for all data types used and needed, respectively. The upper-right quadrant represents data highly used and needed; the upper-left, data highly needed but not highly used; the bottom-right, data highly used but not highly needed; and the bottom-left, data neither highly used nor highly needed.

The discussion emerged at the workshop and the survey results indicate that pandemic modeling and response have in many ways become a hybrid enterprise, in which non-traditional data often complement and augment more traditional (if somewhat slower) sources. This trend, already evident before COVID-19, was only strengthened and solidified during the pandemic.

**Difficulties and Challenges**

The pandemic revealed some significant challenges left unsolved. Both the convened experts and the survey aimed to dig deeper into understanding the nature of these challenges and how they might be overcome, to do better in future health emergencies.

One of the overarching findings of the survey is that the use of data[49] to inform the response to COVID-19 and other public health emergencies can be broken down into two main challenges: access, in which data is identified, sourced, and digested; and translation, in which data is used for policymaking and meaningful interventions. We call these first-mile and last-mile bottlenecks (see **Figure 4** for a resume of the two). Following the survey results, we employed this division to shape the structure of the workshop sessions. First-mile bottlenecks often get the lion's share of attention; however, last-mile shortcomings are equally significant in the way they impede the conversion of insights into timely action. In other words, access without translation is potentially as limiting as a simple lack of access.

**1. "First-mile" Challenges:** First-mile hurdles can arise at any stage of a public health crisis. They refer to difficulties responders face in identifying and accessing relevant data streams[50]. Obstacles can include opaque ownership and licensing, a lack of existing data-sharing agreements, missing metadata, and quality issues. Technical fragmentation—including non-standard formats and incompatible APIs—is another impediment. Two key obstacles highlighted by workshop participants and survey respondents included a lack of standardization and a shortage of incentives for sharing, leading to what the experts labeled reduced "data solidarity"[51].

The survey quantifies some of these pain points. According to experts, roughly 66 % of all datasets—traditional and non-traditional—suffered at least one access problem (see **S1 Fig**). The proportion of issues linked to data sharing reluctance in the context of NTD was double that of traditional data (30% vs. 15%) (see **S1 Fig**). Furthermore, only 55% of non-traditional streams were publicly available or obtainable via government agreements, as compared to more than 83 % for traditional data (see **S2 Fig**). In addition, data quality problems were also prevalent. Despite being rare, preparation and processing issues exist across both traditional and non-traditional data (12% vs. 10%) (see **S3 Fig**). Only 4% of the datasets that participants reported (both traditional and non-traditional) had no quality problems; perhaps somewhat surprisingly, machine readability[52] issues were far more prevalent with traditional data as compared to non-traditional data (11% vs. 2%). These quality issues may stem from the fact that individuals and organizations are less familiar with non-traditional data requirements than traditional data.

**2. "Last-mile" Challenges:** Last-mile issues surface at the opposite end of the data-to-policy pipeline. They hamper the process of turning analysis into actionable intelligence and meaningful interventions. Once again, there are multiple drivers of these limitations, including gaps in data and technical literacy among officials, weak institutional bridges between scientists and policymakers, and slow feedback loops.

Mismatches between the types of data available and requirements for policymaking can also play a role. As we have elsewhere argued[53], taking a questions-based approach to evidence-based interventions can help align data with the most pressing and relevant policy issues.

At a broad level, the survey suggests a degree of success on this front. As noted above, 79% of respondents stated that their analyses fed directly into public-health policy. Nonetheless, a further dive into the survey results suggests a number of obstacles that prevented data from actually being usable (see **S4 Fig**). Besides data availability, these include limitations in the timeliness and granularity of data, issues with linkage[54], comparability, and biases**.** Only 10% of respondents reported that they could use all the data they needed. This was particularly relevant for non-traditional data (e.g., contact tracing[55], mobility, and contact data) and reflects a broader problem with data availability, which remains a significant barrier for data usage in policy making (see **S4 Fig**).

**Survey findings and workshop proposed actions**

In this section, we outline key proposed actions in response to ten lessons learned that emerged from the workshop and survey addressing first and last-mile challenges.

| First Mile | | Last Mile | |
|---|---|---|---|
|  | The need for standardization |  | Accelerating data-driven decision-making |
|  | Mapping non-traditional data |  | Collating evidence from the crisis |
|  | Identify data needs for each pandemic stage |  | Strengthen literacy and capacity |
|  | Foster data solidarity |  | Sustainable funding |
|  | Institutionalize data stewards |  | Streamline collaboration |

Figure 3: Proposed actions addressing first and last-mile challenges

**1. The need for standardization:** Participants at workshop stressed how the experience of reusing non-traditional data during the COVID-19 pandemic revealed widespread inconsistencies in data formats, metadata, and quality metrics. These made it difficult to integrate and analyze data across sources and impeded timely insights. The expert workshop emphasized the urgent need to establish common standards not only for the data itself, but also for metadata structures, reproducibility, and quality indicators. In particular, the lack of standardized approaches[56,57] to processing origin-destination matrices[58] in mobility data emerged as a significant obstacle for modeling[59] and understanding epidemics[60]. One suggestion was the establishment of a multi-sectoral task force to create a common set of questions and methodological standards, enabling smoother collaboration before the next crisis.

One virtuous example is the effort carried out by the MNOdata4OS and Multi-MNO project[61]. This initiative, developed by the European Statistical System[62], aims to create a consistent methodology for processing Mobile Network Operator (MNO) data for official statistics throughout Europe. The project also includes defined quality standards and an open-source reference implementation of the outlined methodology. Notably, this ongoing work seeks to transition MNO data usage from Experimental Statistics[63] to official statistics by addressing challenges of data privacy, creating consistent legislation for long-term data access partnerships, and implementing standardized methodologies.

**2. Accelerating evidence-based decision-making:** Advanced analytics mean little if they do not feed into policy; as we have seen, the last-mile challenges remain significant. The workshop therefore endorsed three "bridges" that could help advance evidence-based policy: (i) Fusion Centers[64], to merge and make actionable data streams; (ii) Decision Accelerator Labs[65], to prototype interventions and iterate on policy; and (iii) a network of Scientific Ambassadors trained to translate data insights into clear-language policy, and more generally to bridge science and practice. Together, these mechanisms can help move from insight to action more quickly. The group also recommended building a cross-border European community of epidemic modelers, so as to expand access to evidence and ensure that no region is left behind in future emergencies.

**3. Mapping non-traditional data:** We observed during COVID-19, as with prior and subsequent public emergencies, that many stakeholders often were not aware of non-traditional data. Participants at the workshop stressed this as a critical barrier and emphasized the need for a "data map", i.e., a comprehensive data inventory describing what datasets exist and rating them for accessibility, relevance, and quality. They also underlined the need for regular audits and harmonized metadata templates so that such an inventory would be searchable and valuable across a range of use cases.

The EU Health Data Space (EHDS)[66], currently collecting health data only, should begin to absorb such demand by building common standards, practices, and governance for health data re-use in research, innovation, and policy-making, integrating NTDs that are essential for fast public health response to health crises. Adopted in January 2025, the EHDS proposes HealthData@EU[67], an ecosystem of national platforms that establishes standards for data quality and use and facilitates the annual publication of data catalogues to find and use common data types across borders.

**4. Collating evidence from the crisis:** Participants at the workshop stressed that, while hundreds of initiatives during COVID-19 used mobility, payments, or social-media signals[3], lessons learned remain scattered and hard to generalize. The workshop highlighted the need for more research on how non-traditional data is used, during which phases, and with what impact. Such a repository would include not only project descriptions but also tags, pipeline schemes, and metadata that allow sorting by disease stage, data type, and outcome at the level of intervention effectiveness based on these data. In this way, future public health professionals and researchers could quickly identify what worked, avoid redundancy, and inform fast and effective interventions while avoiding previous mistakes.

**5. Identify data needs for each pandemic stage:** We learned how identifying and aligning data needs with evolving goals at different stages of a pandemic is essential. Workshop participants highlighted the value of tabletop simulation exercises as a practical tool to explore evolving data requirements. The panel of experts suggested tabletop simulations to provide simulated environments[38] to explore hypothetical pandemic scenarios, revealing dynamic data requirements at various stages of public health emergencies. In doing so, they help policymakers and other stakeholders assess preparedness and response strategies, as well as identify critical data needs and data streams ahead of time.

**6. Strengthen literacy and capacity:** We have noted a number of last-mile hurdles that limit policymakers' capacity to transform non-traditional data into actionable insights. In addition to those already listed, the challenges also include a significant skills gap among public health officials and other stakeholders. To address this, the expert panel called for investments in both analytical training (e.g., interpreting mobility or social media data) and "question literacy"[53]. Question literacy is an essential but under-appreciated skill. It aligns data supply with data needs, and can help prioritize scarce public resources and policymaker bandwidth (e.g., by minimizing time spent searching for or analyzing unhelpful or unnecessary data).

**7. Foster data solidarity**: A recurring theme of the workshop was the need to foster a culture of "data solidarity", i.e., the idea that data holders, especially in the private sector, should commit to sharing data responsibly and proactively during public health emergencies. This requires not only ethical and legal frameworks, but also the cultivation of mutual trust and shared purpose among stakeholders and across sectors. Practical manifestations of data solidarity could include pre-negotiated commitment frameworks, increased transparency around how data will be used, and the development of social licenses[68] through public engagement. In addition, data collaboratives[69], i.e., expert groups based on public-private collaborations exchanging data to help solve public problems, and similar bodies have proven effective in promoting trust and data sharing. Across all these mechanisms, it is essential to establish incentives for sharing; these could include naming and framing schemes and access to aggregated insights for strong sharers.

**8. Institutionalize data stewards:** The expert panel stressed that even when the incentives and structures for data solidarity exist, coordinating actual data sharing is a complex task, especially in the midst of a public health emergency. The workshop suggested the creation of dedicated "data stewards"[70] to help institutionalize and facilitate this process. Data stewards are individuals or groups responsible for building partnerships, conducting audits and risk assessments, facilitating internal coordination, and communicating externally with stakeholders. Their role has proven especially effective in nurturing and sustaining data collaboratives[69]. By making this a formal, resourced role, organizations can ensure they are better prepared to access and share data when a crisis occurs.

**9. Sustainable funding**:  Long-term, sustainable funding and investment are critical to promoting data sharing and effective use of non-traditional data. One of the clearest lessons from the COVID-19 crisis is that ad hoc or one-off funding mechanisms are inadequate, and that long-term capacity requires sustained investment in infrastructure, human capital, and technology. Toward this end, the expert group recommended establishing a dedicated Data Fund, i.e., public-private partnerships supporting data availability for evidence-based policies. The Fund should be governed transparently, include mechanisms for rapid-response grants, and build in accountability through independent oversight.

**10. Streamline collaboration through advanced technologies:** The expert group highlighted how policy, norms, and society are vital parts of any framework to promote the use of non-traditional data. It should not be overlooked, however, that advanced technologies themselves can play a transformative role, especially in encouraging more data sharing. Participants suggested considering technologies such as privacy-enhancing tools[71], Edge Computing[72], Data Sandboxes[73], Data Pods[74], and synthetic data[75], including social contacts[76] and mobility[58] beyond electronic health records. Many of these underlie the emerging concept of "data spaces"—structured, rule-based environments in which data holders retain control over how their data is shared and used[77]. By investing in these technologies, governments and public health institutions can significantly strengthen their ability to collaborate while still maintaining public trust and protecting privacy and other rights.

**Conclusions**

The ten lessons outlined above, as well as the survey results, point to a set of broader priorities for policymaking. In this final section, we distill the experience of COVID-19 into three overarching recommendations that can inform not only future responses to public emergencies, but indeed, steps that governments and other stakeholders need to take today to prepare for the next pandemic. Taken together, these recommendations emphasize the need for sustained investment, institutional change, and a more collaborative, integrated approach to the reuse of non-traditional data in public health emergencies.

**Institutionalize Readiness:** Across all themes, a central theme emerges: success during the next crisis will not be determined by improvisation or timely responses, but by the foundations we lay today. Being better prepared means building durable technical, organizational, and human capacity. Governments and other stakeholders must work today to formalize roles like data stewards, establish data sharing agreements, and invest in tools (e.g., Fusion Centers[64], Decision Accelerator Labs[65], simulation exercises) that can be activated quickly when they are needed. Readiness should be seen as a continuous state, supported by ongoing trust building and community involvement. Some specific actions include:

- Establish and institutionalize data stewards, empowered with formal mandates.

- Create and fund Fusion Centers and Decision Accelerator Labs to prototype, evaluate, and scale interventions based on non-traditional (and traditional) data.
- Conduct regular tabletop simulation exercises involving all stakeholders to enhance readiness and identify weak points or challenges.

**Build Data Preparedness:** Institutional readiness must be accompanied by data preparedness. Once again, many of the most pressing challenges during COVID-19 stemmed from the simple fact that public health representatives were not ready: data was inaccessible, fragmented, and its nature and location poorly understood. Building data preparedness requires developing standardized approaches to how data is structured, stored, and analyzed— especially for non-traditional data such as mobility traces or social media signals. Importantly, there is a need to map what data exists and where it resides. Specific actions include:

- Develop data inventories and maps to identify non-traditional datasets, assess their accessibility and quality, and ensure metadata is harmonized across sources.
- Launch a multi-sectoral task force to set and maintain data standards, including protocols for metadata.
- Create an evidence repository to collate and tag successful (and unsuccessful) uses of non-traditional data during COVID-19 to increase data-literacy.

**Create a Trusted and Collaborative Ecology:** Data does not exist in a vacuum. Its value is realized through relationships among sectors, stakeholders, and, importantly, between those who hold data and those who need it. Trust is the foundation of these relationships, and must be deliberately cultivated. This includes establishing clear ethical guidelines for data use, but also developing institutional intermediaries such as data stewards and data collaboratives that can manage partnerships and facilitate reuse. An observatory can play a particularly critical role, helping to monitor how non-traditional data is used in real time, documenting best practices, and ensuring that insights are shared across borders and sectors. Building this ecology will require a cultural shift, from data hoarding to data solidarity.
Specific actions include:

- Develop mutual commitment frameworks between private data holders and public entities that may use data to clarify terms of emergency data sharing ahead of crises.
- Create an observatory for non-traditional data use that tracks deployment, use, and disseminates real-time insights and lessons learned.
- Encourage and scale data collaboratives with clear governance rules and mechanisms for shared infrastructure and trust-building across sectors and stakeholders.

Together, these three recommendations offer a framework, not a set of independent guidelines or policies. None of these can succeed in isolation, but if pursued together, they can create a virtuous cycle of preparedness, insight, and action.

Crucially, these steps are not only relevant to future pandemics. They can also inform how societies respond to a range of public crises, from climate shocks to humanitarian emergencies. Effective responses to each of these increasingly rely on prepared, tested, informed, and inclusive decision-making.

COVID-19 was an indisputable calamity. Nonetheless, it was also a moment of extraordinary innovation, experimentation, and most of all, learning. If we absorb its lessons and begin now to build the foundations we lacked, then we may yet emerge from this crisis more resilient and better prepared next time. We may not be able to prevent the next crisis—but we can certainly improve our response, and in so doing, potentially mitigate the scope for another calamity.

**Limitations**
The survey was circulated among experts in Europe. Experiences of experts from the rest of the world may highlight further lessons learned and suggest further recommendations.
Despite circulating the survey through general channels such as the ECDC Modeling Hubs mailing list, we among the ESCAPE project modelers and their networks and to the workshop participants, selection bias could still affect participants' responses as having used NTD during the pandemic could have shaped the decision of participants to the survey.
Our findings and recommendations remain general and generalizable to other emergencies and diseases, as none of the data discussed or actions proposed at the workshop are specific to the modeling of and response to COVID-19.

**Acknowledgments**
Workshop participants, namely:
André Peralta Santos Health Directorate Portugal, Acting Deputy Director, General for Health
Arnon Vandenbergh, King Baudouin Foundation, Data Scientist
Arnout Desmet, Blauwe Cluster Sustainability Advisor
Brecht Ingelbeen, Institute of Tropical Medicine, Antwerp Researcher
Casey Weston, LinkedIn - Data for Impact, Program Senior Manager
Christophe Fraser, University of Oxford - Nuffield, Dept. of Medicine, Professor of Epidemiology
Ciro Cattuto, ISI Foundation, Scientific Director
Constantino Caetano, Portuguese National Health Institute Doutor Ricardo Jorge, Ph.D. Student
Daniela Paolotti, ISI Foundation, Research Leader
David Dab, Microsoft, National Technology Officer
Denis Xavier Renaud, Orange Head of Orange Flux Vision R&D
Emma Hodcroft, Swiss Tropical and Public Health Institute, Assistant Prof. and leader of the Epidemiology and Virus Evolution group

Esteban Moro, Northeastern University, Professor in Physics
Francesco Parino, INSERM, Post-Doc Researcher
Helena Coning, Mastercard, Chief Privacy & Data Responsibility Officer
Hien Vu, CEPS, Associate Researcher
Ilaria del Seppia, European Commission - HERA.03, Policy Officer
Ingmar Weber, Saarland University Alexander von Humboldt, Professor for AI and
Chair for Societal Computing
Irma Varela-Lasheras, Portuguese National Health Institute Doutor Ricardo Jorge,
Post-Doc Researcher
Jasmin Menten, The Data Tank, Executive Assistant
Javier de la Cueva, IE University, Lawyer, Researcher and Lecturer
Jean-Michel Contet, Orange
Jill Falman, Meta, Data for Good Program Manager
Jonny Shipp, The Data Tank, Consultant
Jose Javier Ramasco, Spanish National Research Council, Researcher
Julia Ebert, Vodafone Institute, Senior Research Manager
Kaylin Bolt, Public Health Seattle King County, Social Research Scientist
Kyriaki Kalimeri, ISI Foundation, Researcher
Laura McGorman, Meta Data for Good, Director
Laura Sandor, The Data Tank, Programme Manager
Leo Ferres, Universidad del Desarrollo, Associate Professor of Computer and Data
Science
Linus Bengtsson, Flowminder, Chair of the Board
Lisa Hermans, Hasselt University, Research Manager
Luca Ferretti, University of Oxford - Pandemic Sciences Institute, Career Development
Fellow
Luca Pappalardo, Italian National Research Council, Senior Researcher
Lukas Adomavicius, Centre for Information Policy Leadership (CIPL), Privacy Analyst
Manuel Garcia Herranz, UNICEF, Chief Scientist
Manuel Ribeiro, University of Lisbon, Assistant Researcher
Maria Loureiro, European Commission - HERA.02, Intelligence Analyst - Health
Threats
Mattia Mazzoli, ISI Foundation, PostDoc Researcher
Mohammad Reza, Rahmanian Haghighi University of Antwerp, PhD Candidate
Moiz Shaikh, The Data Tank, Programme Manager
Natascha Gerlach, Centre for Information Policy Leadership (CIPL), Director of Privacy
& Data Policy
Niel Hens, Hasselt University, Director & University of Antwerp, Scientific chair holder
Olena Snidalova, Vodafone Institute, Senior Research Manager
Paul Theyskens, MyData Global, Board Member
Paula Patricio, NOVA School of Science and Technology, Associate Professor
Paula Vasconcelos, Support Unit of National Health Authority and the Emergency
Management in Public Health, Public Health Emergencies Coordination Support
Operations Centre - Portugal
Paulina Behluli, The Data Tank, Project Associate
Pietro Coletti, Hasselt University/JRC, PostDoc Researcher
Polen Türkmen, The Data Tank, Project Lead

Richard Benjamins, OdiseIA, CEO & Co-Founder
Rik Gosselink KU Leuven Ethics Advisor
Slim Turki, Open Data & Data Ecosystems, Senior Researcher
Sofie Bekaert, King Baudouin Foundation, Senior Program Manager
Sonia Namorado, Portuguese National Health Institute Doutor Ricardo Jorge, Researcher
Stefaan Verhulst, The Data Tank, Co-Founder and Principal Scientific Advisor
Stefan Thurner, Complexity Science Hub Vienna, President
Takahiro Yabe, New York University, Assistant Professor

**Funding Statement**

This project was supported by the ESCAPE project (101095619), funded by the European Union. Views and opinions expressed are however those of the author(s) only and do not necessarily reflect those of the European Union or European Health and Digital Executive Agency (HADEA). Neither the European Union nor the granting authority can be held responsible for them. M.M, P.T., L.F., K.K., C.C., D.P., S.V., acknowledge support from the Lagrange Project of the ISI Foundation, funded by Fondazione CRT. L.F. acknowledges support from the Fondo de Investigación y Desarrollo en Salud, Fonis, Project SA24I0124.

**Authors' Contributions**

NH, CC, DP and SV acquired the funding and conceptualized the work. IVL, SN, CPC, AL and DP created the survey. MM, PT, KK, LF, CC, DP and SV organized the workshop. IVL analyzed the data and visualized the results. MM, IVL and SV wrote the original draft. All authors contributed to reviewing and editing the final version.

**Conflicts of Interest**

None declared.

**Abbreviations**

AIS - Automatic identification systems
EC – European Commission
EHDS - EU Health Data Space
MNO – Mobile network operators
NTD – Non traditional data
WBE – Water based epidemiology
XDR – eXtended detail records

Supplementary material

**Survey on data readiness and availability during the COVID-19 pandemic.**

**Supplementary methods**

This study employed a cross-sectional online survey design to assess data availability, use, and unmet needs for the modeling community in epidemiology during the COVID-19 pandemic.

The survey was directed to research groups or individual researchers focused on epidemiological modeling, primarily in the European context. To this end, the survey was disseminated through the ECDC Modeling Hubs mailing list, among modelers in the ESCAPE project and their network, as well as in a high-level expert workshop on the use of non-traditional data during pandemics in Brussels, which included representatives of academia, government agencies, and the private sector. The survey was available between March and October 2024, in the REDCap and LimeSurvey platforms.

The survey had the following specific objectives:

1. To gain a better understanding of the types of data used for modeling during the pandemic, their purpose, and their strengths and limitations, with a particular focus on non-traditional data.
2. To gain a better understanding of the unmet data needs of the modeling community and the underlying causes.

The survey was completely anonymous and consisted of 7 main questions. There were two questions: one regarding "data types used" and another regarding "data types not used but needed." The options for data types were numerous, and there was free text space for other types not provided. These two questions had several subquestions, allowing us to collect details for each specific data type reported (e.g., details about data, access, and quality, or reasons that prevented the use of data)

The survey was divided into 3 sections.

- The first one aimed to understand the type of research questions and modeling approaches that were used by the research groups/researchers who answered the survey.

- The second section aimed to identify (1) what data was available and used during the pandemic by these research groups; (2) how data was accessed and what problems were encountered in accessing the data; and (3) what quality problems were encountered when using the data.

- The third section aimed to understand what data were needed but not used by these research groups and whether these unmet needs were the result of data not being available or the result of data readiness problems (quality and access issues).

Throughout the survey, the data were categorized into traditional epidemiological data (TD; e.g., tests, hospitalizations, deaths, vaccination, etc.) and non-traditional epidemiological data (NTD), defined as data types not collected with epidemiological purposes and/or relatively new data types (e.g., mobility, contacts, wastewater data, etc.). Thus, in addition to each individual data type, we could have a broader view of traditional vs. non-traditional data.

**Supplementary results**

In total, 29 research groups/researchers responded to the survey. Mathematical/mechanistic models were the most common method used. The most common research questions were related to the evaluation of control measures, description and short-term prediction of the pandemic, and monitoring of its impact.

The analysis of the survey showed that TD types were the most widely used and available (TD was used by 96% of the respondents and accounts for 67% of the responses)(Fig.3). TD was mainly publicly available or accessed through agreement with governments (83% of responses) (S2 Fig.). However, different types of access and quality problems persist in about 66% and 96% of these cases, respectively (S2 and S3 Fig.). NTD types were commonly used, although less often than TD types (90% of respondents and 33% of responses) (Fig.3). NTD was also often publicly available or accessed through agreements with governments (55% of responses), although agreements with private institutions and direct access were also very common (S2 Fig.). Hospitalizations and mobility data were the most commonly used TD and NTD types, respectively (used by 82% and 72% of respondents). As for TD, for NTD, different problems with access and quality were also present in around 65% and 96% of cases, respectively (S1 and S3 Fig.).

Regarding the unmet data needs, NTD data types account for more than 72% of the needed but not accessible data (Fig.3). In particular, contact tracing, contacts, and socio-economic data were highly needed but not accessible. Mobility data, although often used, was still highly needed. Regarding the reasons behind these unmet data needs, for both TD and NTD data, unavailability (either total or in adequate time) prevented use in around 60% of the cases (S4 Fig.). Other relevant obstacles for data use were biases (for TD) or temporal and spatial granularity, as well as temporal availability (for NTD).

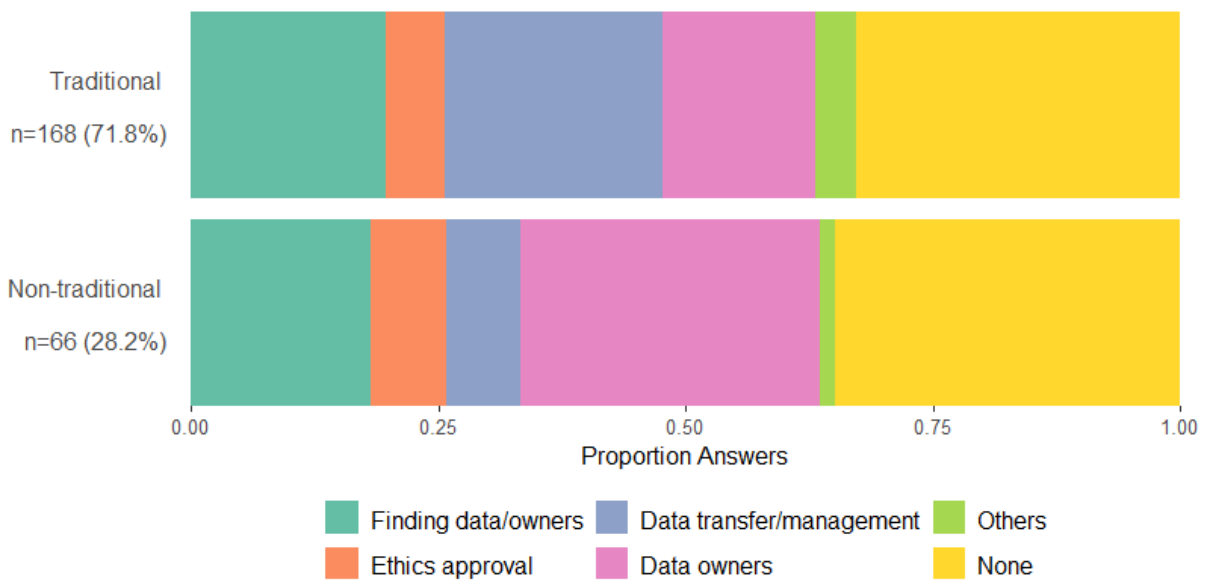

**S1 Fig.** Problems regarding data access for each of the data types used by respondents, represented by the proportion of answers for each type of data access problem aggregated in TD and NTD. Note that for simplicity, several categories are shown aggregated ("Data owners" aggregates the following categories: data owners reluctant/unwilling to share for privacy, national security, academic competition, and commercial reasons). The total number of answers for TD and NTD, and the percentage of the total number of answers are also shown (note that for each data type, respondents could choose more than one answer).

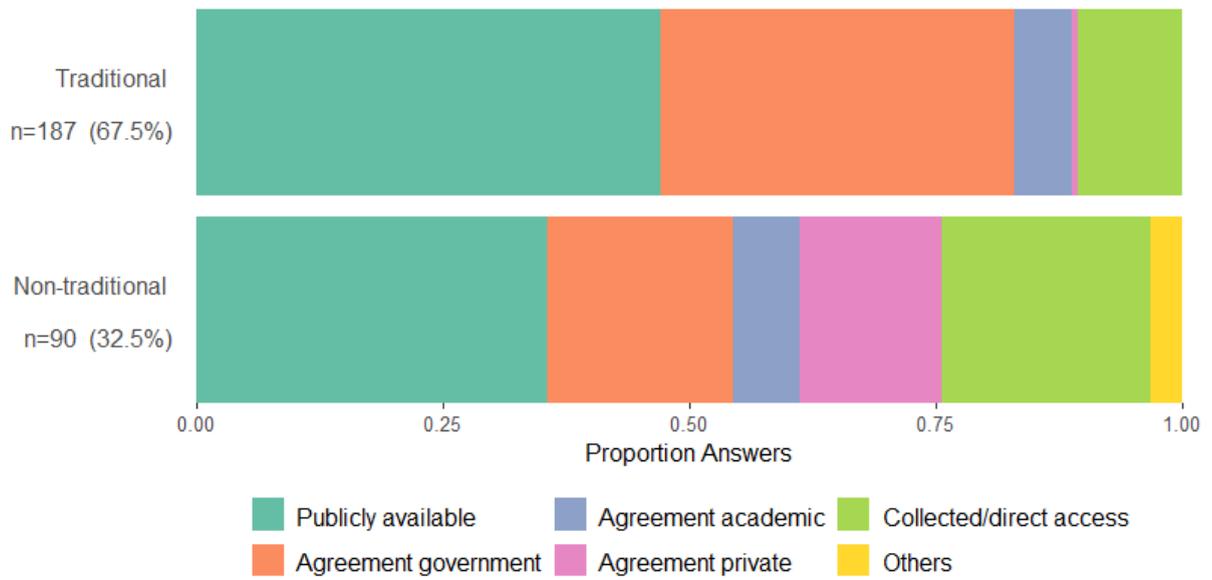

**S2 Fig.** Type of access for each of the data types used by respondents, represented by the proportion of answers for each type of data access aggregated in TD and NTD. The total number of answers for TD and NTD, and the percentage of the total number of answers are also shown (note that for each data type, respondents could choose more than one answer).

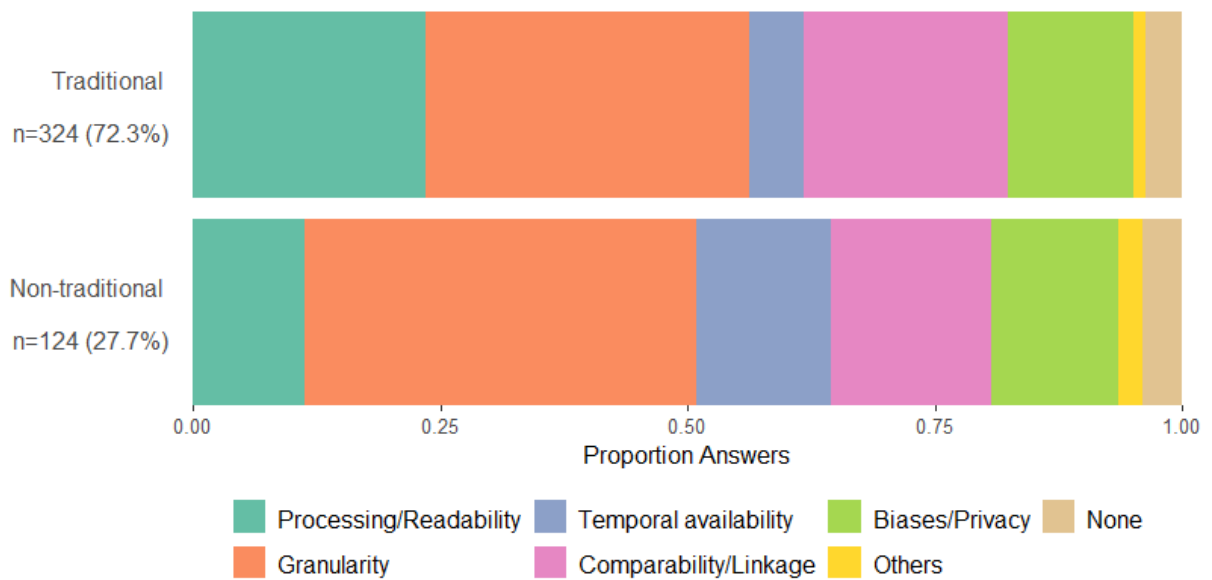

**S3 Fig.** Problems regarding data quality for each of the data types used by respondents, represented by the proportion of answers for each type of data quality problem aggregated in TD and NTD. Note that for simplicity, several categories are shown aggregated (e.g., "Processing/Readability" aggregates problems with preparation and processing, and with machine readability; "Granularity" aggregates problems with temporal, spatial, and demographical/medical granularity). The total number of answers for TD and NTD, and the percentage of the total number of answers are also shown (note that for each data type, respondents could choose more than one answer).

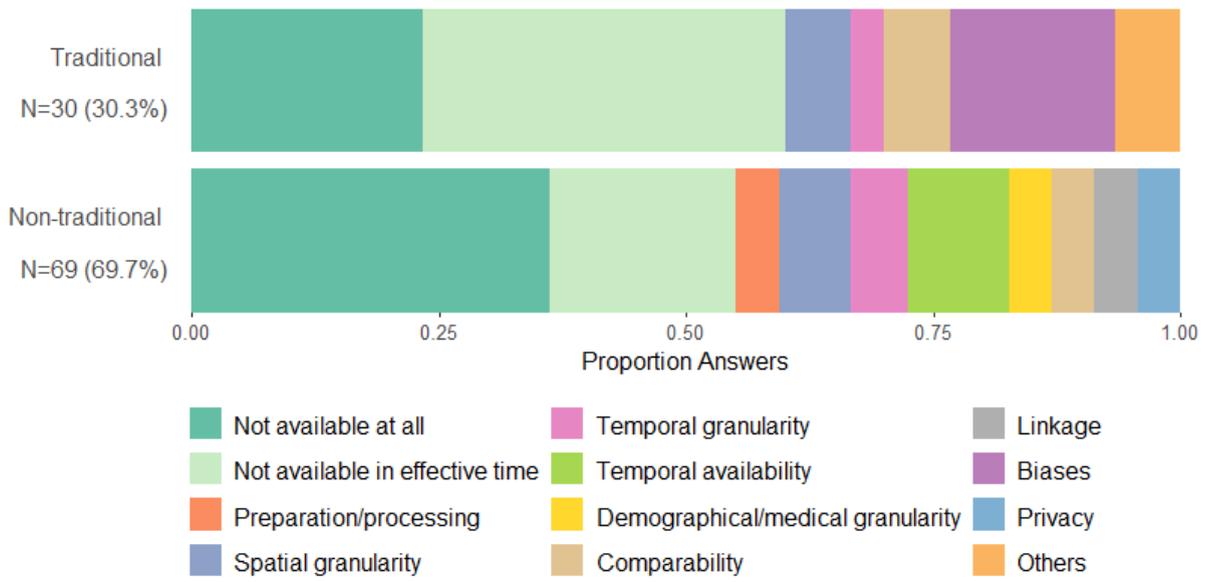

**S4 Fig.** Reasons that prevented the use of data for each data type not used reported by respondents, represented by the proportion of answers for each of the reasons listed, aggregated in TD and NTD. The total number of answers for TD and NTD, and the percentage of the total number of answers are also shown (note that for each data type, respondents could choose more than one answer).